# Optimization of Dense Wavelength Division Multiplexing demultiplexer with 25GHz uniform channel spacing

Venkatachalam Rajarajan Balaji[a], Mahalingam Murugan [b], Savarimuthu Robinson[c]

[a]Department of ECE, St.Joseph's Institute of Technology, Research scholar, Anna University, Chennai, India

[b]Department of ECE, Valliammai Engineering College , Anna University, Chennai, India

[c]Department of ECE, Mount Zion College of Engineering and Technology , Anna University,Pudukkottai , India

Address all correspondence to Balaji.V R, Department of ECE, and St.Joseph's Institute of Technology,AnnaUniversity,Chennai,India.Email: photonicsdemux@gmail.com

**Abstract:** In this paper ,we propose a four channel Dense Wavelength Division Multiplexing demultiplexer with two dimensional photonic crystal square resonant cavity that fulfill the ITU-T recommendation  of G.694.1 DWDM systems.DWDM demultiplexer consists of a  waveguide and Microscopic Square Resonant (MSR) cavity to enable filtering of the desired  wavelength. The MSR cavity design has inner rods, outer rods and coupling rods. As the radius of the inner rod in MSR cavity changes, the cavity has the ability to filter different ITU.T G.6941 standard wavelengths like 1555.3 nm, 1555.5 nm, 1555.7 nm, and 1555.9 nm with 0.2 nm / 25 GHz channel spacing. From the simulation of various wavelengths, helps achievement of the quality factor of 8000, uniform spectral linewidth of 0.2 nm, transmission efficiency of 100 %, crosstalk of -42 dB and footprint is about 395 μm$^2$.

**1. Introduction**

Optical fibers transmit light, between the two ends of fiber, which permits transmission with a longer distance at data rates higher than that existing communication network. The Network is highly expensive when each user associated with a single fiber of obliges the existing million users due to the presence of millions of fiber and becoming complicated. The solution for this issue lies in a single mode fiber simultaneously for various customers.

The single mode optical fiber (SMF) utilizes WDM and DWDM procedures, for transferring multiple light wave with distinct wavelengths inside the SMF .The transfer of multiple light wavelengths with SMF in the other end of the network uses a optical demultiplexer for splitting the resonant wavelengths for each client. There are two types of optical demultiplexer devices namely, passive demultiplexers and active demultiplexers. The Passive demultiplexer has frequency filters [1, 2], prisms [3], diffraction gratings [4]. On the other hand, the active demultiplexer has a combination of passive components and tunable detectors. The existing technology limits low normalized transmission power, high crosstalk, low quality factor, and scale of centimeters, resulting in inability to adapt to Photonic Integrated Circuits [PICs].

PICs dilute the existing technology drawbacks and scale to micrometer demultiplexer devices. In PICs, a Photonic crystals (PCs) reduces the device scale in micrometer and the able to control hundreds of channel within same scale. The PCs together with Photonic Band Gap (PBG) control the propagation of light [5, 6]. No light with different wavelengths can propagate through the structure, the structure region is PBG.

The PBGs allow wavelength to propagate, and create defects in the periodic structure. In general, PCs defect formed in two ways; line defects and point defects, the line defect is formed with removal or change in structural -parameters (lattice constant, refractive index, radius of the rods) of the entire row of rods in the design structure. The point defect is created due to alterations the structure parameters of the single rod or removal thereof. Designing PC devices requires introduction of both defects in the proposed structure. The PC is used for designing a wavelength division demultiplexer [7] , Beam Splitter [8], Optical Switches [9] , Ring Resonators [10] ,Photonic Sensors[11] and etc.

WDM is categorized into Coarse WDM (CWDM) and Dense WDM (DWDM). The CWDM type provides up to eight wavelengths with 20 nm channel spacing, whereas the DWDM type provides up to two fifty six wavelengths with 0.2 nm / 25 GHz channel spacing. In recent years, researchers have been showing greater attention towards the DWDM device due to its unique features. Generally, DWDM technology accomplishes dynamic usage of bandwidth, low attenuation components of single mode fibers, it uses multiple wavelength as carriers and concedes them for simultaneous transmission in the fiber simultaneously [12], with ability to support up to two fifty six wavelengths with 0.2 nm / 25 GHZ of spectral linewidth. The aforementioned features provide the solution for problems experienced by service providers such as capacity crisis, high speed, high flexibility, and long reach. The main function of the optical demultiplexer is to separate the wavelength for coupling to individual fibers.

Literature survey , brings to light reports from the 2D PC based demultiplexer for CWDM and DWDM system using line defects [22] and ring resonators [13-16] ( The shape of the ring resonator in created through introduction of line and point defects). Considering the fruitful benefits in ITU.T. G.694.1 DWDM system, authors consider DWDM system here. The DWDM demultiplexer using 2D PC is made by T shape structure with line defects resonant cavity [17],P shaped single resonant cavity with different rod radius to drop different wavelength [18],multi T-shaped structure with line//point defects [19-21],hetero structure resonant cavity [22] and X ring cavity [23]. Literature survey identifies the design of demultiplexer using cavity of different shapes. However, the transmission efficiency, crosstalk, Q factor do not meet the requirements. Hence, in this paper, a new square ring resonant cavity based demultiplexer is proposed. It is designed for enhancing the aforementioned functional parameters.

In this paper, a four-channel DWDM demultiplexer is proposed and designed .The demultiplexer utilizes new MSR cavity modeled with outer rods, inner rods, coupling rods. The designed MSR cavity can filter the desired wavelength by changing the inner rods and coupling rods, while keeping other parameters such as like outer rod, lattice constant, refractive index are constant. The proposed model MSR cavity is designed with square size, which reduces the scattering losses, which, in turn, enhances coupling efficiency. Hence our proposed model highlights with certain features like low crosstalk (> - 40 dB), high transmission efficiency (> 98), high quality factor (narrow line width), footprint. These are not found in the earlier works. The paper begins with a review of two-cavity filter analysis and then expands to four channel DWDM systems.

This paper is organized as follows: Section 2, discusses the PBG and its range of proposed structure before introducing the defects. Section 3, proposes the design of two cavity structure and four-cavity structure .The simulations results and discussion are reported in Section 4, Section 5 concludes the paper

**2. Photonic crystal Geometry:**

The proposed DWDM filter utilizes 2-Dimensional square lattice photonic crystal for better horizontal confinement of light. The filter is built with 31×47 rods in X and Z direction for effective coupling of modes, each silicon rod having a radius R=115 nm, where lattice constant a = 520 nm and a refractive index of 3.5 which is embedded in air.

The theoretical analysis of 2D PC for Plane Wave Expansion method (PWE) and Finite Difference Time Domain method (FDTD) is studied. The PWE method calculates the propagation of modes and the PBG of periodic and non-periodic structures. The propagation modes of electromagnetic waves in photonic crystal analysis satisfies Maxwell's equations [24, 25]

$$\nabla \times \left( \frac{1}{\varepsilon(r)} \nabla \times E(\text{r}) \right) = \frac{\omega^2}{c^2} E(\text{r}) \quad (1)$$

Where, $\varepsilon(r)$ represents the dielectric function, '$\omega$' the angular frequency, $E(\text{r})$ represents the electric field of periodic structure and 'c' is the speed of the light. Eq. (1) 2D determines the PC band structure. The FDTD method simulates the propagation of electromagnetic waves inside the PCs. [27].

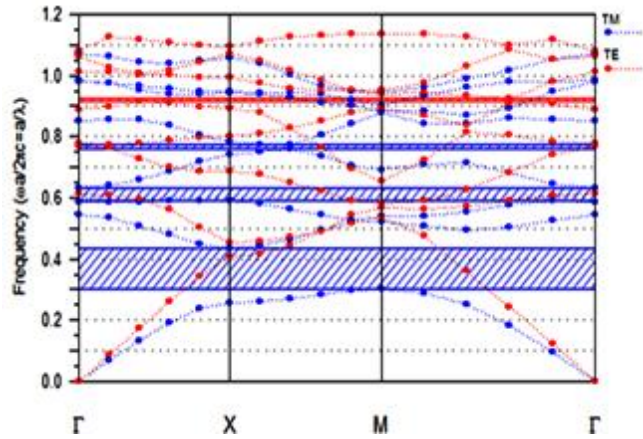

**Fig. 1.** Band diagram for 31*47 PC structures before the inducing the defects.

Fig.1. Shows the band diagram with 31×47 PC structure before inducing the defects. It has three TM PBGs and a TE PBG in band diagram as shown in Fig.1.The type of PBG, Normalized frequency and its wavelength range are listed in Table 1.The table, shows first TM PBG wavelength between 1168 nm -1730 nm as meant for low loss communication, since the frequency lies in third window of optical communication, which is considered for our work. The entire simulation is performed with TM mode where the electric field is perpendicular to the rod axis. The proposed crystal band structure design uses a two/four cavity DWDM filter

**Table 1**

Photonic band gap, normalized frequency and its corresponding wavelength of the of the proposed structure.

| S.NO | PBG | Normalized frequency($a/\lambda$) | Wavelength(nm) |
|------|-----|-----------------------------------|----------------|
| 1 | TM PBG | 0.3-0.445 | 1168-1730 |
|   |        | 0.588-0.6455 | 805-884 |
|   |        | 0.768-0.778 | 668-677 |
| 2 | TE PBG | 0.923-0.934 | 556-563 |

**3. PC Based Two/Four channel filter design**

The PC of two/four-channel cavity design schematic proposes a band gap structure. The Fig. 2. Shows the two-channel cavity filter schematic.

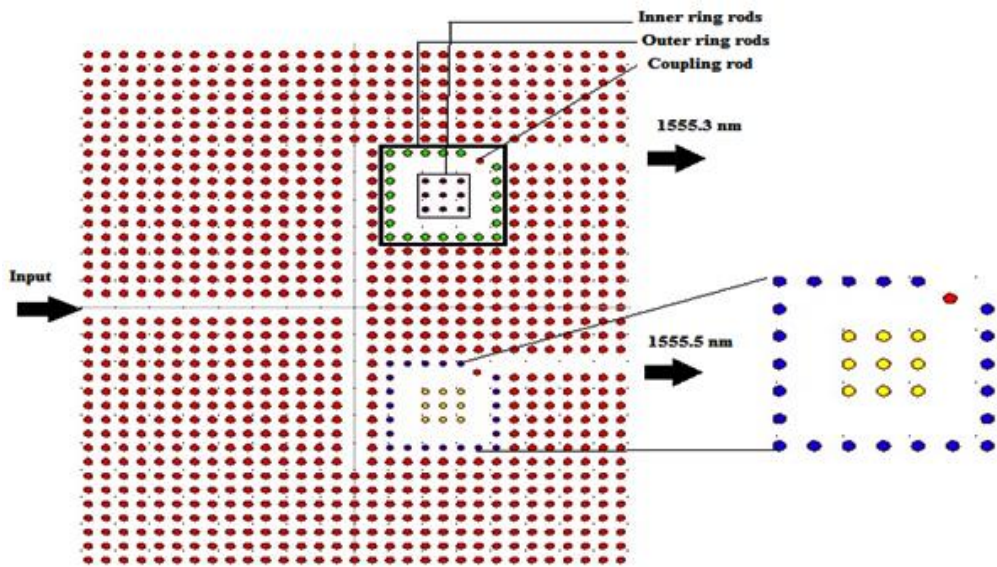

**Fig. 2.** Schematic view of 2D PC based square ring resonant cavity

The proposed MSR cavity filter consists of inner rods, outer rods, coupling rods and waveguide. The waveguide is formed with line defects where line/point defects are created for getting a square shape of resonant cavity. In resonant cavity, inner rods placed inside the cavity, rods placed outside the cavity are outer rods. The coupling rod is located at the right corner of the ring resonator for coupling the signal from square resonant cavity to waveguide.

The T shape waveguide utilizes two channels DWDM filter in the crystal. The radius of outer and inner, coupling rods to design demultiplexer is list in Table 2.

**Table 2**

Radius of the outer, inner, and coupling rods of the proposed demultiplexer

| Channels | Radius of Outer rods | Radius of Inner rods | Radius of Coupling Rods |
| --- | --- | --- | --- |
| $\lambda_1$ -1555.3 nm | 115 nm | 76 nm | 95 nm |
| $\lambda_2$ -1555.5 nm | 115 nm | 76.4 nm | 95 nm |

The radii of the rods determine optimized simulations under different conditions, like characteristics of rods, lattice constant, and refractive index. The Perfect Matched Layer Absorbing Boundary Conditions (PML ABC) is utilized for avoiding anti reflection of electromagnetic waves for all the frequencies and angle of incidence. It is formulated with the help of the Maxwell's equations. . The proposed crystal structure utilizes PML ABC (PML Absorbing Boundary Conditions) for optimizing the boundary region without reflection of electromagnetic waves [28, 29].

The absorption rate is higher in the PML ABC structure, whereas other boundary conditions show more reflections during simulations. The power monitor is positioned at the end of the waveguide to enable obtaining a normalized output transmission.

Fig.3. shows the normalized output spectra with a two channel filter. The resonant wavelength transmission efficiency and quality factor of the proposed filter are 1555.3 nm, 1555.5 nm, 100 % and 7777, 7778, respectively. The quality factor can be calculated by using [26]

$$Q = \frac{\lambda_r}{\Delta \lambda} \qquad (2)$$

Where, $\lambda_r$ is the resonant wavelength, $\Delta\lambda$ is the spectral line width or full width as half-maximum. This Eq. (2) helps calculation of the Q factor. The spectral line width is 0.2 nm obtained at 1555.3 nm, sufficient to drop a channel for DWDM. As the demultiplexing function itself utilizes a filter, two channel filters is extended to four channels for dropping four distinct channels.

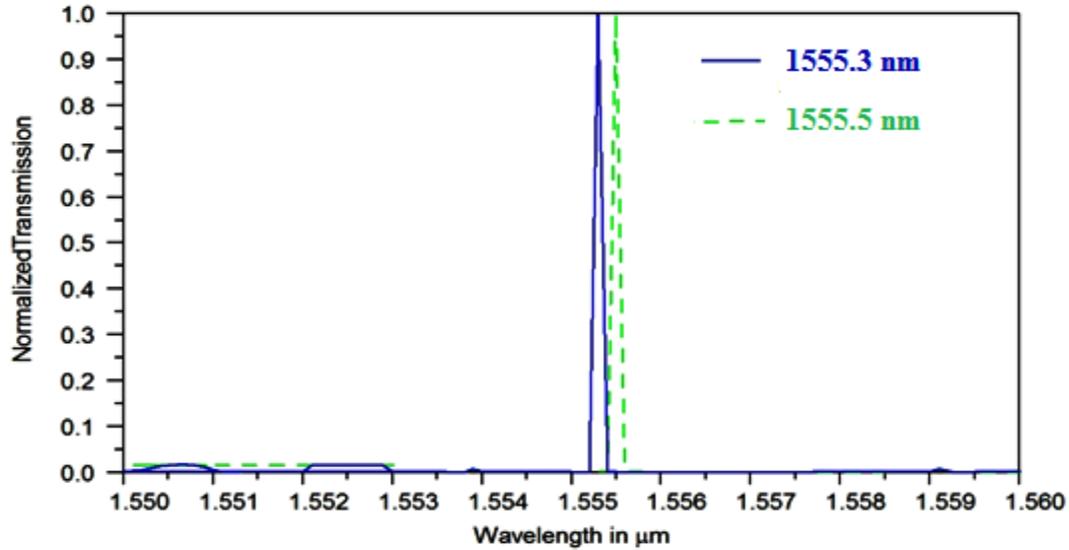

**Fig. 3.** Normalized output spectrum of the proposed square ring resonant cavity for single channel.

The demultiplexing function of the filter is to drop the desired wavelength for distinct channels. The filter extends to a four-channel port from the two-channel port for dropping four desired wavelengths. The four-channel port demultiplexer schematic representation is shown in Fig. 4**.**The demultiplexer includes four resonant cavities, with each resonant cavity being responsible for dropping the desired wavelengths. The desired wavelengths are dropped with a distinct coupling rod radius and radii of inner rods as like a filter. Other parameters such as like outer rod, lattice constant, and refractive index are constant.

The filter works with increasing the inner rods radius, that placed inside the microcavity. The dielectric strength of the rod increases based on the radius of the inner rods. Increasing the dielectric strength causes to shift the resonance in higher resonant wavelength. For dropping the four distinct wavelength size of the internal rods are Ri=76 nm,76.4 nm.76.8 nm and 77.2 nm repectively.The designed demultiplexer selects the radii of the inner rods with optimized simulation under different conditions, like characteristics of rods, lattice constant, and refractive index. The demultiplexer is designed with four ports to drop the four desired wavelengths.

Table 3, shows the detailed structure parameters for each individual distinct channel. The channels are designed to drop the desired wavelength with high transmission efficiency, high Q, and narrow linewidth

For separating the demultiplexer ports, in ports 1 & 2 and 3 & 4 ports separate with 4 row of rods. The separation rods count between 2 & 3 ports increase more than four row of rods due to bend curve nature. The nature of the bend curve helps the photons to the desired coupling cavity, which fails when the separating rods are smaller in number. This entry of photons to the 2 and 3 port with a less number rods causes these deviation from the coupling concept theory.

**Table 3** Radius of the Outer, Inner and Coupling rods of the proposed demultiplexer.

| Channels | Radius of Outer Rods | Radius of Inner Rods | Radius of Coupling Rods |
|---|---|---|---|
| $\lambda_1$ -1555.3 nm | 115 nm | 76 nm | 95 nm |
| $\lambda_2$ -1555.5 nm | 115 nm | 76.4 nm | 95 nm |
| $\lambda_3$ -1555.7 nm | 115 nm | 76 .8 nm | 95 nm |
| $\lambda_4$ -1555.9 nm | 115 nm | 77.2 nm | 95 nm |

The results show the four-channel demultiplexer using square ring resonant cavity. It is composed of a T shaped waveguide and four MSR cavities where each resonator is responsible for dropping different channels. The counter propagation modes are suppressed with placing the coupling rod at the right corner of the ring resonator, which reduces the back reflection, which in turn enhances the constructive interference .The result, validates the designed demultiplexer with transmission efficiency, high Q factor, narrrow spectral linewidth, low crosstalk.

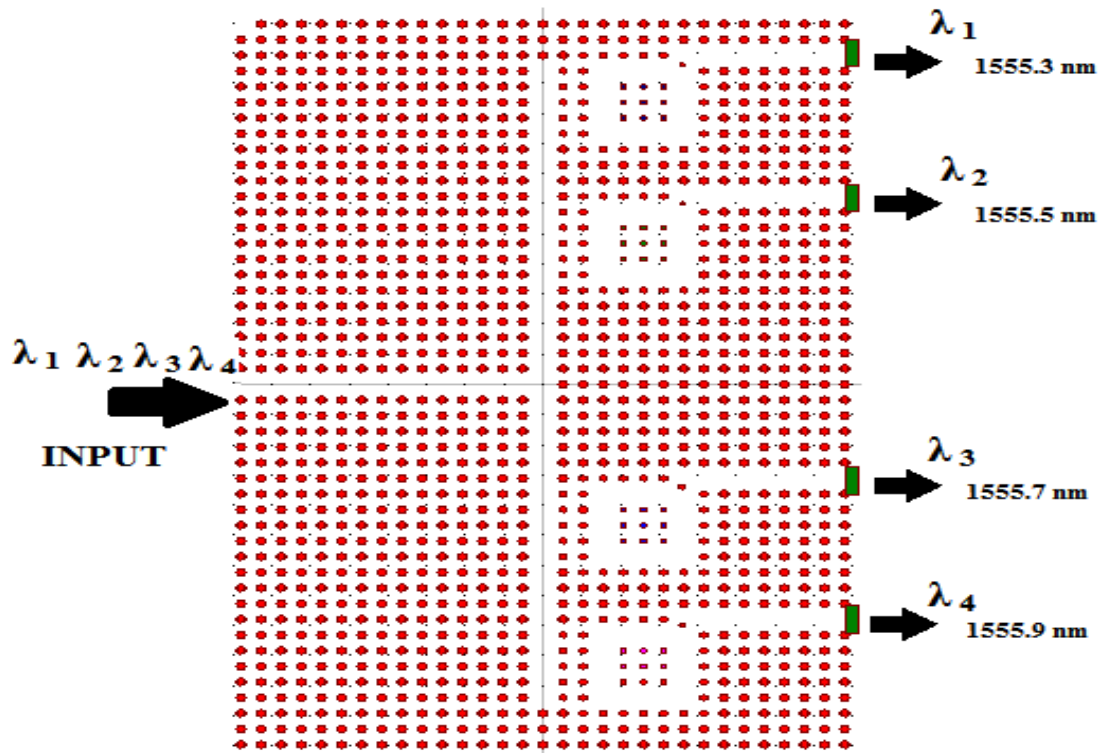

**Fig. 4.** Schematic diagram of proposed four-port demultiplexer using PC square resonant cavity

The 3D view of proposed four channel demultiplexer is depicted in Fig.5. The size of the demultiplexer is 16.2 μm × 24.4 μm which is very small; Hence it could be easily deployed in PICs.

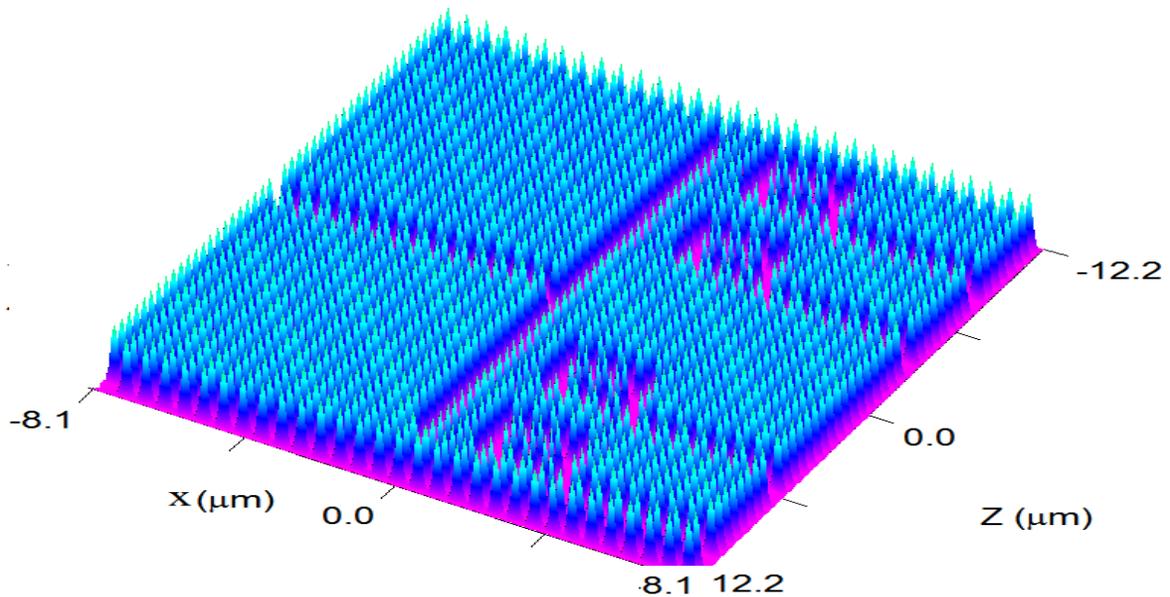

**Fig. 5.** 3D view of proposed four-channel demultipler.

## 4. Results and Discussion

The proposed four-channel demultiplexer utilizes the spatial pulse source as spatial gaussian shape. The pulse source at the input port of the waveguide is in nanometer wavelength, which demultiplexes with resonant cavity to measure the power at each corresponding output port. The simulation perform using FDTD with PML ABC, the width of PML and PML reflection for the design are considered as 500 nm and $10^{-8}$ respectively. The FDTD grid size in the simulation is maintained at X/20(0.05a=26 nm) to get significant results for the DWDM environment. Maintenance of precise time step during the simulation is significant for getting accurate results. The time step obeys the following condition in the filter

$$\Delta t \leq \frac{1}{c\sqrt{\frac{1}{\Delta X^2} + \frac{1}{\Delta Y^2}}} \qquad (3)$$

In Eq. (3), $\Delta t$ represents the step time, and c represents the speed of light in free space. The filter simulates with increment of 0.0001 for a 3600 min run time for memory structure of 41.8 MB to get high Q factor output.

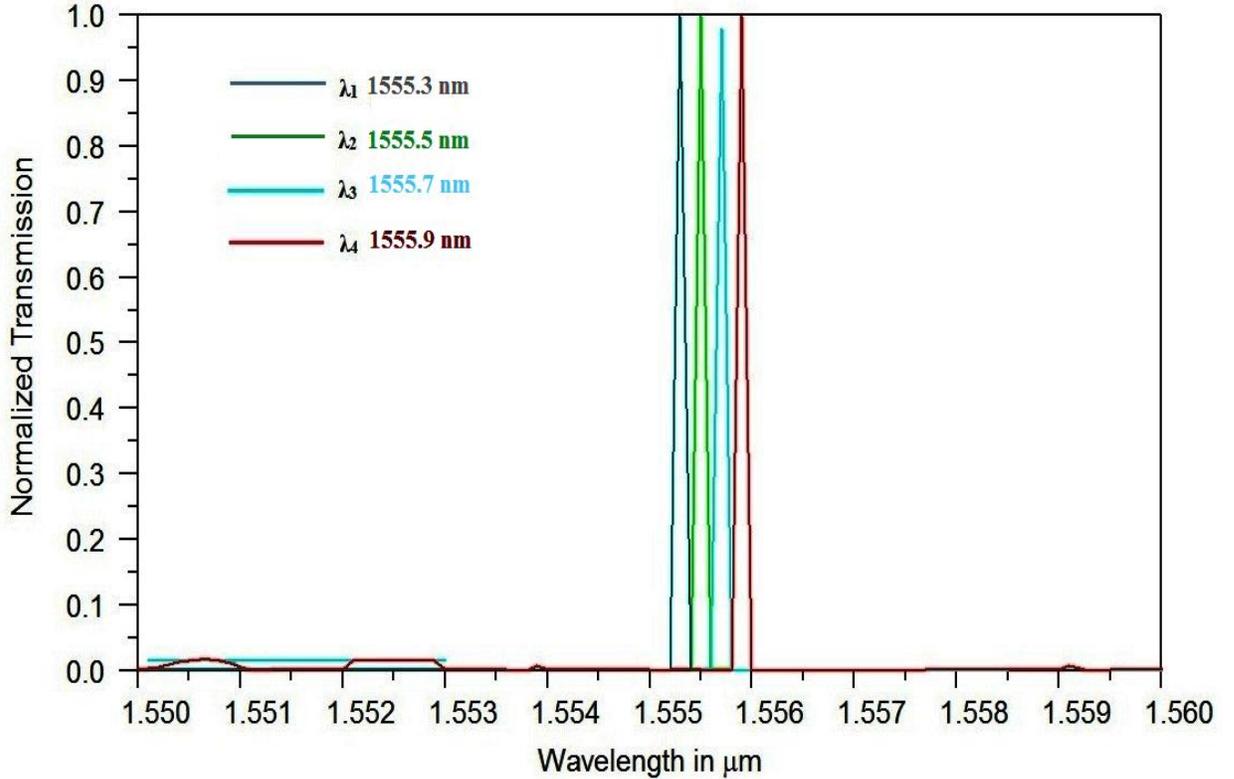

**Fig. 6.** Normalized output spectra of four-channel DWDM demultiplexer.

The output spectra of the proposed four-channel demultiplexer are shown in Figure.6.The wavelengths in the Fig. attain ITU.T. G.694.1 DWDM systems**.** The resonant wavelengths of the DWDM system is observed at 1555.3 nm, 1555.5 nm, 1555.7 nm, and 1555.9 nm that lie between C band of optical window. The C band window is widely preferred to the network due to low loss communications. The simulation shows 100% coupling efficiency, 0.2 nm spectral linewidth, channel spacing, and quality factor is almost equal to 8000.The results meet the requirements of ITU .G 694.1 DWDM systems. The resonant wavelength, spectral line width, output transmission efficiency and quality factor of the proposed demultiplexer depict are depicted in Table 4.

**Table 4**

Resonant wavelength, Transmission efficiency, Spectral linewidth and Q factor of four channel PC based demultiplexer

| Channels | Resonant Wavelength $\lambda_r$ (nm) | Transmission Efficiency | Spectral linewidth $\Delta \lambda_r$ (nm) | Q Factor |
|---|---|---|---|---|
| $\lambda_1$ | 1555.3 | 100 % | 0.2 (**1.5554-1.5552**) | 7777 |
| $\lambda_2$ | 1555.5 | 100 % | 0.2 (**1.5556-1.5554**) | 7778 |
| $\lambda_3$ | 1555.7 | 99 % | 0.2 (**1.5558-1.5556**) | 7779 |
| $\lambda_4$ | 1555.9 | 100 % | 0.2 (**1.556-1.5558**) | 7780 |

The biggest challenge in designing DWDM demultiplexer is to obtain low cross talk. The demultiplexer design is focused on improvement of low crosstalk with constructive interference based MSR cavity. The spectral response of demultiplexer in dB scale is in Fig.7, and is meant for calculation of crosstalk among the channels (Pij). Fig.7, shows variation of the crosstalk of the channels over the range from -27.8 dB to -42 dB, which produces a very small crosstalk, compared with previous works.

The crosstalk between channels is Pij, where i and j denote the channel numbers. For example $P_{13}$ gives the crosstalk between channel 1 and channel 3. The crosstalk among the channels is list in Table 5.

**Table 5**

Crosstalk values (Pij) of proposed four channel PC based demultiplexer (dB)

| Channels(Pij) | λ₁ | λ₂ | λ₃ | λ₄ |
|---|---|---|---|---|
| λ₁ | - | -30.17 | -33.43 | -42 |
| λ₂ | -30.25 | - | -33.1 | -38.8 |
| λ₃ | -33 | -33.43 | - | -33.33 |
| λ₄ | -42 | -39.11 | -33.04 | - |

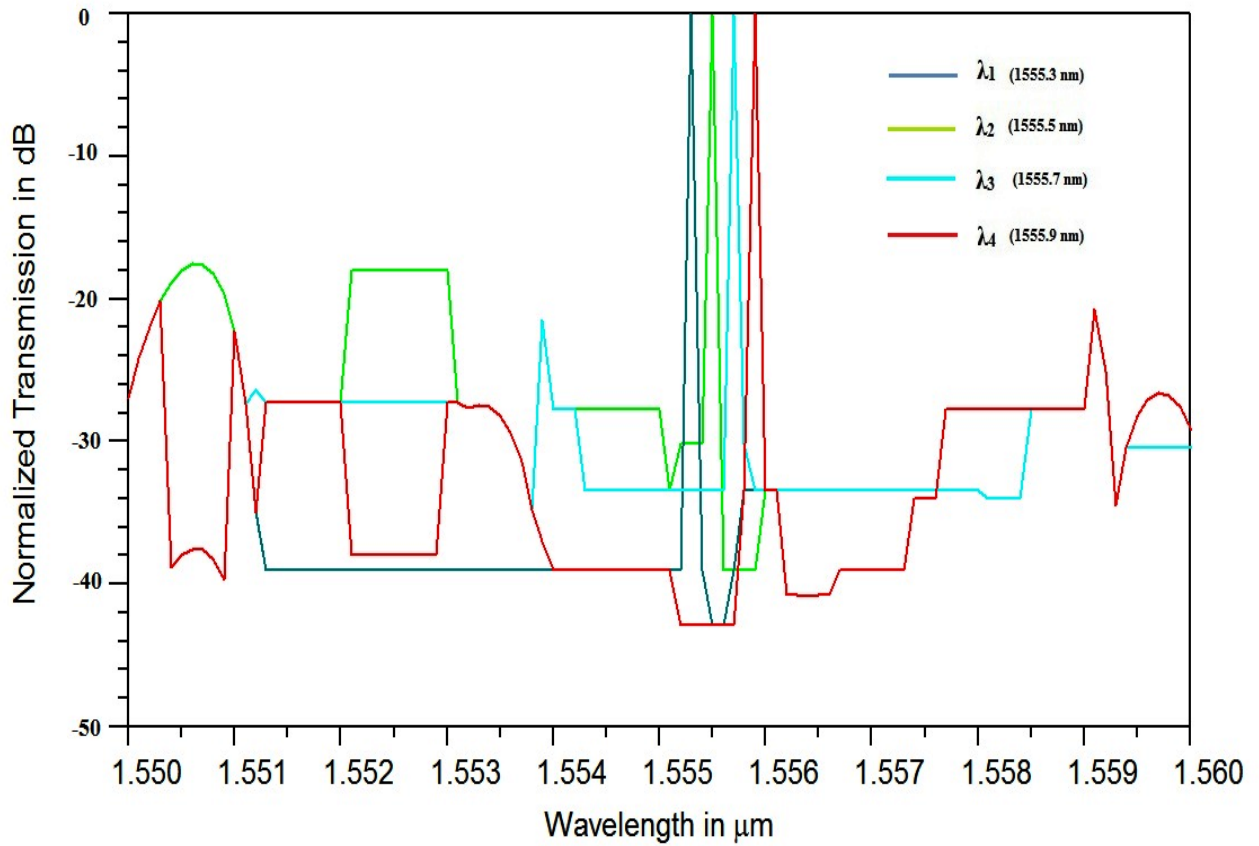

**Fig. 7.** Spectral output for four channel DWDM demultiplexer in dB scale.

The functional characteristics of the proposed four port DWDM demultiplexer is compared with the reported DWDM demultiplexers, which are provided in Table 6 . The Table 6, proves the work of the proposed square resonant cavity as superior to that of the existing DWDM. The simulation of various wavelengths, helps achievement of functional parameters such as quality factor of 8000, spectral linewidth of 0.2 nm, transmission efficiency of 99% - 100 % crosstalk of -42 dB and the device size is about 395 µm$^2$. The results shows the non uniform spectral linewidth found in existing system is replaced to the uniform spectral linewidth for the dropping wavelengths The functional parameters are significantly enhanced with smaller footprint , and hence the proposed demultiplexer can be implemented in the Photonic Integrated Circuits

**Table 6**

The proposed Microscopic Square resonator based DWDM demultiplexer is compared with the reported DWDM demultiplexer in the literature.

| Authors/Year | No of Output Ports | Coupling Efficiency (%) | | Q Factor | | Crosstalk(dB) | | Foot Print (µm)$^2$ | Spectral Linewidth (nm) |
|---|---|---|---|---|---|---|---|---|---|
| | | Min | Max | Min | Max | Min | Max | | |
| Rostami et al. [17]/2009 | 4 | 42.5 | 86.5 | 3006 | 3912.5 | -30.00 | -14.2 | 536 | 0.4 |
| M.Djavid et al.[15]/2012 | 3 | 82 | 91 | NA | NA | NA | NA | NA | NA |
| Mohammad reza rakhnshani [22]/2013 | 6 | 80 | 90 | 2000 | 2319 | -34 | -23 | NA | 2 |
| Hamed Alipoureta et.al [23]/2013 | 4 | 45 | 63 | 561 | 1954 | -23.70 | -7.5 | 422.4 | 2.8 |
| Mohammed ali Mansouri et al.[13]/2013 | 3 | 80 | 96 | 390 | 891 | -29 | NA | 317 | 3.8 |
| Abbasgholi et al.[20][2014] | 2 | 52.61 | 63.38 | 5264 | 7900 | -24.59 | -19.6 | NA | 0.2 |
| Nikhil Deep et al.[21]/2014 | 4 | 40 | 60 | 7795 | 7807 | NA | NA | NA | 0.2 |
| Farhad Mehdizadel [29]/2015 | 8 | 94 | 98 | 1723 | 3842 | -40 | -11.2 | 495 | 1 |
| **This work** | **4** | **99** | **100** | **7775** | **7790** | **-42** | **-30** | **395** | **0.2** |

**5. Conclusion:**

The proposed four channel Dense Wavelength Division Multiplexing demultiplexer with two dimensional photonic crystal square resonant cavity fulfills the ITU-T recommendation of G.694.1 DWDM systems. The novelty in the proposed DWDM is the dropping the desired wavelength by altering the radius of the inner rods, and which is position in the MSR cavity. The spectral response of the proposed DWDM performs about 100 % of transmission efficiency, -42dB crosstalk with the quality factor of 8000. Further, the channel spacing and spectral linewidth between the channels are 0.2 nm/ 25 GHz and 0.2 nm, respectively. The proposed PC based demultiplexer is excellent in fulfilling the requirements of ITU.T.G.6941 DWDM system and size is very small about 395 μm$^2$, it could be incorporate for integrated optics. From the results, the existing DWDM system with non-uniform spectral linewidth for drop wavelengths replace to the uniform spectral linewidth. The crosstalks of proposed DWDM demultiplexer improve about -42 dB compare to previous works.

**Acknowledgements**:
    The author Balaji would like to acknowledge Dr.N.R.Shanker, Manager, Research and Development, Chase Technologies, Chennai for his support at different times.

**Conflict of Interests**
There is no conflict of interests regarding the publication of this paper.

**Caption List**

**Fig.1.** Band diagram for 31*47 PC structures before the introduction of defects

**Fig.2.** Schematic view of 2D PC based square ring resonant cavity

**Fig.3.** Normalized output spectrum of the proposed square ring resonant cavity for single channel

**Fig.4.** Schematic diagram of proposed four-port demultiplexer using PC square resonant cavity

**Fig.5.** 3D view of proposed four-channel demultiplexer. .

**Fig.6.** Normalized output spectra of a four-channel DWDM demultiplexer.

**Fig.7.** Spectral output for the four channel DWDM demultiplexer in dB scale.

**Table 1** Photonic band gap, normalized frequency and its corresponding wavelength of the of the proposed structure

**Table 2** Radius of the outer, inner, and coupling rods of the proposed demultiplexer

**Table 3** Radius of the outer, inner and coupling rods of the proposed demultiplexer

**Table 4**.Resonant wavelength, Transmission efficiency, Spectral linewidth, and Q factor of four channel PC based demultiplexer

**Table 5** Crosstalk values (Pij) of proposed four channel PC based demultiplexer

**Table 6** The proposed Square ring resonator based DWDM demultiplexer is compared with the reported DWDM demultiplexer in the literature.